\documentclass[prb,twocolumn,showpacs,preprintnumbers,amsmath,amssymb,notitlepage]{revtex4-1}
\usepackage{graphicx}
\usepackage{ulem}
\UseRawInputEncoding

\begin{document}

\title{ Bose-Einstein condensations of magnons in quantum magnets with spin-orbit coupling in a Zeeman field }
\author{ Fadi Sun$^{1,2}$ and Jinwu Ye$^{1,2,3}$   }
\affiliation{
$^{1}$ The School of Science, Great Bay University, Dongguan, Guangdong, 523000, China \\
$^{2}$ Institute for Theoretical Sciences, Westlake University, Hangzhou 310024, Zhejiang, China  \\
$^{3}$  Department of Physics and Astronomy, Mississippi State University, MS 39762, USA     }
\date{\today }

\begin{abstract}
We study the response of a quantum magnet with spin-orbit coupling (SOC)  to a Zeeman field by constructing effective actions and performing
Renormalization Group (RG) analysis. There are several novel classes of quantum phase transitions at a low
$ h_{c1} $ and an upper critical field $ h_{c2} $ driven by magnon condensations at commensurate (C-)
or in-commensurate (IC-) momenta $ 0 < k_0 < \pi $.
The intermediate IC- Skyrmion crystal (IC-SkX) phase is controlled by  a line of fixed points in the RG flows
labeled by $ k_0 $.
We derive the relations between the quantum spin and the order parameters of the effective actions
which determine the spin-orbital structures of the IC-SkX phase.
We also analyze the operator contents near $ h_{c1} $ and $ h_{c2} $ which determine the exotic excitation spectra inside the IC-SkX.
The intrinsic differences between the magnon condensations at the C- and IC- momenta are explored.
The two critical fields $ h_{c1} < h_{c2} $ and the intermediate IC-SkX phase could be a generic feature
to any quantum magnets with SOC in a Zeeman field.
Experimental implications to some materials or cold atom systems with SOC in a Zeeman field are presented.
\end{abstract}

\maketitle

\section{ Introduction }
 In 1938, The Bose-Einstein condensation (BEC) was first observed in the superfluid state of $ ^4 He $ which is a strongly interacting system \cite{london}.
 In 1995, it was also observed in a dilute gas of alkali atoms which is a weakly interacting system \cite{cold1,cold2}.
In the  Bose-Einstein condensation (BEC),
the global $ U(1)_c $ symmetry corresponding to the boson number conservation is spontaneously broken, it leads to the Goldstone mode with a linear dispersion. It was known
that the magnon condensation tuned by a Zeeman field in a quantum magnet \cite{bec1,bec2,bec3,z2,frusrev,rev1,rev2}
can be mapped to the BEC with the spin $ U(1)_s $ symmetry
mimicking the charge $ U(1)_c $ symmetry of the bosons.
Here, inspired by recent many experimental studies on the response to a Zeeman field of some chiral magnets \cite{sky4}
and 4d or 5d Kitaev materials \cite{halfinteger,unquantized} with strong spin-orbit couplings (SOC),
we study the magnon condensation in a quantum magnet with SOC in a Zeeman field and
find it leads to dramatically new phenomena outlined in the abstract and Fig.1.

The  system is the interacting spinor bosons at integer fillings hopping in a square lattice in the presence of non-Abelian gauge fields
studied in \cite{rh}. In the strong coupling limit, it leads to the spin $ S=N/2 $ Rotated Ferromagnetic Heisenberg model (RFHM)
which is a new class of quantum spin models to describe quantum magnetisms in cold atom systems or some materials with strong SOC.
There is an exact $ U(1)_{soc} $ symmetry along the anisotropic line $ (\alpha=\pi/2, 0 < \beta < \pi/2) $ of the 2d SOC.
Along the line, we identified a new spin-orbital entangled commensurate ground state: the Y-x state.
It supports not only commensurate magnons (C$_0$,C$_{\pi}$),
but also a new gapped elementary excitation: in-commensurate magnon ( IC- ).
The existence of the C-IC  above a commensurate phase is a salient feature of the RFHM.
They indicate the short-ranged in-commensurate order embedded
in a long-range ordered commensurate ground state.
The IC- magnons may become the seeds to drive possible new classes of quantum C-IC transitions under various external probes.
In \cite{rhh}, by performing the microscopic spin-wave expansion (SWE),  we explored the effects of
an external longitudinal Zeeman field $H$ applied to the RFHM Eq.\ref{rhh} along the anisotropic SOC line
which keeps the $ U(1)_{soc} $ symmetry.
However, the microscopic SWE approach used in \cite{rhh} is essentially a semi-classical approach. It may not apply to a small
quantum spin $ S $ in real materials, will also break down near all the quantum phase transitions in Fig.1.
A complete independent symmetry based phenomenological
effective action is needed to study the nature of these novel quantum phase transitions.

 In this work, starting from general symmetry principle, we construct various effective actions, then
 perform Renormalization Group (RG) flows and carefully analyze the physically accessible initial conditions to study all the quantum phase transitions in Fig.1. We also identify the relations between the quantum spins in a lattice and the order parameters in the effective actions, which are needed to identify the quantum spin-orbital orders of the phases.
 When away from the quantum critical points, we recover all these quantum phases and their excitations discovered by the microscopic calculations in \cite{rh,rhh},
 Most importantly, we explore the nature of all the quantum phase transitions,
 therefore provide deep insights into the global phase diagram in Fig.1.
 The transition from the Z-x phase to the in-commensurate Skyrmion crystal (IC-SkX) phase  at $ h=h_{c1} $ is in the same universality class as the $ z=2 $ Superfluid (SF)-Mott transition \cite{z2,rev1}.
 In addition to the well known Type-I dangerously irrelevant operator (DIO) \cite{z2,rev1,scaling}, there is also
 a new type-II DIO \cite{type2}  which leads to one exotic Goldstone mode inside the IC-SkX phase near $ h_{c1} $.
 At the Mirror Symmetry (MS) point $ \beta=\pi/4 $ (Fig.1),
 the  Type-II DIO is absent, the exotic Goldstone mode recovers to the conventional one.
 The quantum phase transition(QPT) from the Z-FM to the IC-SkX at $ h=h_{c1} $ is described by a novel two component effective action with
 the dynamic exponent $ z=2 $ and a $ U(1)_{soc} \times U(1)_{ic} $ symmetry where the extra $ U(1)_{ic} $ symmetry comes from the condensations of the magnons at the two IC-momenta.
 It was spontaneously broken down to its diagonal $ [ U(1)_{soc} \times U(1)_{ic}]_D $ leading one Goldstone mode inside the IC-SkX phase.
 By performing Renormalization group flow analysis, we find it is a novel universality class with new operator contents.
 In addition to two Type-I DIOs, there are also two type-II  DIOs
 which lead to one  exotic gapless Goldstone mode and one gapped exotic roton mode inside the IC-SkX phase near $ h_{c2} $.
 At the MS point, the two Type-II DIOs are absent, the exotic Goldstone and roton mode recover to conventional ones;
 a quartic Umklapp term  which breaks the extra $ U(1)_{ic} $ symmetry explicitly to $ Z_4 $ and
 becomes the third Type-I DIO and plays important roles. The RG analysis on the effects of this Umklapp term is performed.
 These results may shed considerable lights on the experimental findings, especially the nature of the intermediate phases
 of some materials with strong SOC such as  the chiral magnets and the Kitaev materials in a Zeeman field.
 Some perspectives are outlined.

\begin{figure}
\includegraphics[width=0.4\textwidth]{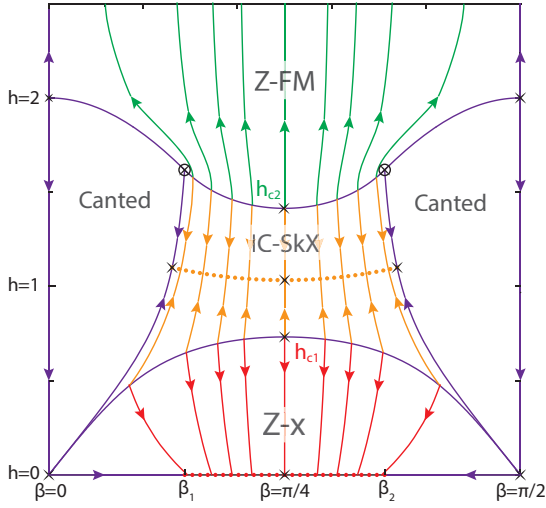}
\caption{
Renormalization group (RG) flow of Eq.\eqref{rhh} which explores many new features of magnetic systems with SOC.
Z-FM is a Ferromagnet along the Zeeman field. IC-SkX is the In-commensurate Skyrmion crystal. Z-x is the Ferromagnet
alternating along the x-bond.
There is a line of  unstable fixed ( or QPT ) points labeled by the BEC condensation C- or IC- momentum $ k_0 $ in Eq.\ref{drop}
and in Eq.\ref{drop2} near $ h_{c1} $ and $ h_{c2} $ respectively. There is
also a line of stable fixed points labeled by the $ 0 \leq k_0 \leq \pi $ inside the IC-SkX phase.
The RG flows just follow the constant contour line of  $ k_0 $.
The new universality classes of all these QPTs and the corresponding operator contents
at or away from the two MS points are analyzed in the text. }
\label{globalphase}
\end{figure}


Here, we focus on the RFHM along the line $ (\alpha=\pi/2, 0<\beta<\pi/2 ) $
in the Zeeman field along the longitudinal $ y $ direction\cite{rhh} ( See also appendix A ).
After rotating spin $Y$  axis to $Z$ axis, it can be written as \cite{twoterms}:
\begin{eqnarray}
	\mathcal{H} & = &  -J\sum_i[\frac{1}{2}(S_i^+S_{i+x}^+ + S_i^-S_{i+x}^-)-S_i^zS_{i+x}^z    \nonumber  \\
     & + &  \frac{1}{2}(e^{i2\beta}S_i^+S_{i+y}^-+e^{-i2\beta}S_i^-S_{i+y}^+) +S_i^zS_{i+y}^z]   \nonumber  \\
    & - &  H \sum_i S_i^z
\label{rhh}
\end{eqnarray}
where $ J >0 $, the Zeeman field $H$ is along the $ z $ direction after the global rotation.


As shown in \cite{rh}, Eq.\ref{rhh} at $ H=0 $ has the translational symmetry,
the time reversal $ {\cal T} $, the three spin-orbital coupled $ Z_2 $
symmetries $ {\cal P}_x, {\cal P}_y, {\cal P}_z $. Most importantly, it also owns
a hidden spin-orbital coupled $ U(1)_{soc} $ symmetry generated by $ U_1(\phi)=e^{ i \phi \sum_{i} (-1)^x S^{z}_i } $.
The $ H $ breaks the $ {\cal T}, {\cal P}_x, {\cal P}_y $ symmetries, but still keeps the translation, $ {\cal P}_z $,
the combinations $ {\cal T}{\cal P}_x, {\cal T}{\cal P}_y $ and the hidden $ U(1)_{soc} $ symmetry.
Under the Mirror transformation $ {\cal T} \cdot {\cal M} $ where
$ {\cal M}=\tilde{\mathbf{S}}_{i} =R(\hat{x},\pi ) R(\hat{z},\pi n_2) \mathbf{S}_{i}$,
$ (\beta, h ) \rightarrow ( \pi/2 - \beta, h ) $. So one only need to focus on the left half of Fig.1.
The mirror center $ \beta=\pi/4 $ respects the MS. In the following, we will take $ 2SJ $ as the energy unit.

\section{  Quantum phase transition at the lower critical field $h_{c1}$ }

The spin wave expansion (SWE) in the Z-x state below $ h_{c1} $ was performed in \cite{rhh}.
Dropping the higher branch $\alpha_\mathbf{k}$ in Eq.\ref{betak}, it is the
$\beta_\mathbf{k}$ magnon condensation at $\mathbf{K}_0=(0,k_0)$ which leads to the QPT from the Z-x state to the IC-SkX at $ h_{c1} $ in the whole range of $ 0 < \beta < \pi/2 $:
\begin{equation}
 \beta_\mathbf{k} =\psi\delta_{\mathbf{k},\mathbf{K}_0},~~~~ \alpha_\mathbf{k} =0
\label{drop}
\end{equation}
where $\mathbf{K}_0=(0,k_0)$ and $\psi$ is a complex order parameter.

 One must use the unitary transformation Eq.\ref{unitaryTrans} to establish the connection between
 the transverse quantum spin on the lattice and the order parameter in the continuum limit:
\begin{align}
    S_{A,i}^+ &=\sqrt{2S} a_i	=c\psi e^{ik_0i_y},
               \nonumber  \\
    S_{B,j}^- & =\sqrt{2S} b_j	=s\psi e^{ik_0j_y}
\label{k0iyjy}
\end{align}
where $c=c_{\mathbf{K}_0}$ and $s=s_{\mathbf{K}_0}$ are evaluated at $\mathbf{K}_0=(0,k_0)$.


The Z-x state spontaneously break the translation along the $x-$ direction by one lattice site
to two lattice site, i.e. $\mathcal{T}_x\to(\mathcal{T}_x)^2$, but still keeps all the other symmetries of the Hamiltonian
 listed below Eq.\ref{rhh}. After incorporating this fact, one can study how $\psi$ transform under the symmetries  of the Hamiltonian,
especially under  $\mathcal{T}_y $ as $ \psi \to e^{ik_0}\psi $,
$ U(1)_{soc} $ as $ \psi \to e^{i\phi_0}\psi $. At the MS point $ \beta=\pi/4 $, under
$\mathcal{T}\circ\mathcal{M} $ as $ \psi \to -\psi^\ast $.
The symmetry analysis suggests the following effective action in the continuum limit with the dynamic exponent $z=2$
\begin{align}
	\mathcal{L}_{h_{c1}}
	&=\psi^\ast\partial_\tau\psi
	+v_x^2|\partial_x\psi|^2
	+v_y^2|\partial_y\psi|^2
	-\mu|\psi|^2
	+U|\psi|^4           \nonumber  \\
	&+iV|\psi|^2\psi^*\partial_y\psi +\cdots
\label{z2single}
\end{align}
where $ \cdots $ include terms such as  $ i\psi^\ast\partial^3_y \psi $ and many other terms which are subleading in the RG sense \cite{factor}.
Our microscopic calculation shows that $ \mu=h-h_{c1} $ which tunes the QPT,
$U>0 $, $ V\propto\sin (2k_0)$ vanishes at $ \beta=\pi/4 $ dictated by the MS.
In fact, its symmetry breaking pattern  \cite{sameasEq10} and associated Goldstone mode is identical to Eq.\ref{coset} to be discussed in Sec.III.

When $ h < h_{c1}, \mu<0$, it is in the Z-x state with $  \langle \psi \rangle =0$.
 Expanding the effective action upto second order
in  $ \psi $ leads to the gapped excitation spectrum
$  \omega_\mathbf{k}
	=-\mu+v_x^2k_x^2+v_y^2k_y^2
$  which matches the results achieved by the microscopic SWE calculation in \cite{rhh}.

When $ h > h_{c1}, \mu>0$, it is in the IC-SkX state with $ \langle \psi \rangle =\sqrt{\rho_0}e^{i\phi_0}$
where $\rho_0=\sqrt{\mu/2U}$ and $\phi_0$ is a arbitrary angle due to U(1)$_\text{soc}$ symmetry.
Plugging the mean field solution into Eq.\ref{k0iyjy}, one obtain
the spin-orbital order of the IC-SkX phase above $ h_{c1} $:
\begin{equation}
   \langle  S_i^+ \rangle =(\sqrt{\rho_0}/2)[c+s+(-1)^{i_x}(c-s)]e^{(-1)^{i_x}i(k_0i_y+\phi_0)}
\label{ordercs}
\end{equation}
   where $ \langle S_i^z \rangle $ can be  fixed by the constraint $|\mathbf{S}_i|^2=S^2$.
   It has a non-vanishing Skyrmion density $Q_{ijk}=\mathbf{S}_i\cdot(\mathbf{S}_j\times\mathbf{S}_k)$
   where $ i, j, k $ are 3 neighbouring lattice sites in a square lattice.
   Well inside the IC-SkX state, one can also extract
   its exotic Goldstone mode due to the $ U(1)_{soc} $ symmetry breaking:
\begin{align}
    \omega_\mathbf{k}
	=\sqrt{4U\rho_0(v_x^2k_x^2+v_y^2k_y^2)}-V\rho_0 k_y
\label{Goldhc1}
\end{align}
  which recovers the conventional Goldstone mode at  the MS point $ \beta=\pi/4 $ where  $ V=0 $.

  In the following, we use the Wilsonian momentum shell method to derive the RG flow near $ h_{c1} $ and $ h_{c2} $ at or away from
the Mirror symmetric point in Fig.1. This method is the quickest one to achieve the RG flow at one -loop order.
However, it is not practical when there is a gauge field, because it is very difficult to keep gauge invariance even at one loop order \cite{RGSF}. It can not be pushed into two-loops either which can only be achieved by Field theory method.
Fortunately, one-loop order is enough to capture the physics in the present context.
We first study the single component
case near $ h_{c1} $ to set-up the scheme and notations, then investigate the more interesting two component
cases near $ h_{c2} $.

\subsection{The RG near $ h_{c1} $ in Fig.1: operator contents at  the Mirror symmetric ( MS ) point  }

The effective action describing  the Magnon BEC in the presence of SOC
near $ h=h_{c1} $ at the Mirror symmetric point $ \beta= \pi/4 $ in Fig.1 is:
\begin{align}
	S_{h_{c1}} & =\int d\tau d^dr
	[\psi^*(r,\tau)\partial_\tau\psi(r,\tau)
	+\frac{\hbar^2}{2m}|\nabla\psi(r,\tau)|^2    \nonumber   \\
	& -\mu|\psi(r,\tau)|^2	+u|\psi(r,\tau)|^4+ \cdots],
\end{align}
where $\psi$ is the complex scalar field. After adding back the type-II dangerously irrelevant operator (DIO) $ V $ in Eq.\ref{z2single},
it describes the QPT near $ h_{c1} $ away from the mirror symmetric point in Fig.1.
As shown in Eq.\ref{z2} in the appendix B, it also precisely describes the Magnon BEC of the AFM in a uniform Zeeman field.

\begin{figure}
\includegraphics[width=0.45\textwidth]{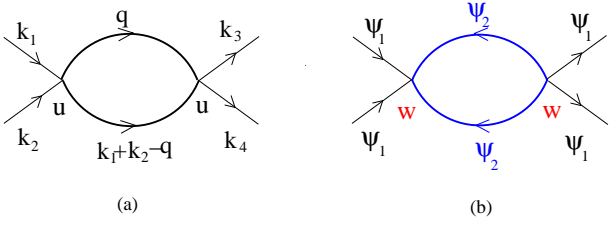}
\caption{ The renormalization  of the self-interaction $ u $. We indicate the 3-momentum $ k=(\vec{k}, \omega ) $.
The black line means $ \psi_1 $,  the green line means $ \psi_2 $. The internal line is on the 2-momentum shell
$ \Lambda e^{-\delta l} < q < \Lambda $. (a) due to $ u^2 $   (b) due to $ w^2 $.  }
\label{oneloopu}
\end{figure}

The major simplification of $ z=2 $  over the relativistic case $ z=1 $ is that
when performing RG in the $ z=2 $ theory, one must separate the frequency
$ \omega $ from the momentum $ \vec{k} $,
then the frequency integral $ \int \frac{d \omega}{2 \pi } $ will kill many Feymann diagrams
which will otherwise make a contribution in a relativistic QFT. Fig.\ref{oneloopu}a is the only surviving diagram.
Similar simplifications also apply to the two component cases to be discussed in Sec.III and Sec.IV.

The Wilsonian RG flow equation at one 1-loop is given by Fig.\ref{oneloopu}a:
\begin{eqnarray}
	\partial_l\mu&= & 2\mu  \nonumber\\
	\partial_l u &= & \epsilon u - c u^2
\label{single}
\end{eqnarray}
where $\epsilon=2-d$ and $c=K_d\Lambda^{d-2}2m/\hbar$ with  $K_d=\frac{S_d}{(2\pi)^d}=\frac{2}{(4\pi)^{d/2}\Gamma(d/2)}$.
In fact we expect that Eq.\ref{single} is exact to any loop order due to the exact $ U(1)_{soc} $ symmetry and $ z=2 $.
Our field theory RG analysis \cite{RGSF} on $ u $  confirms this expectation upto two loops.

In the present context of Fig.1, $ u $ is marginally irrelevant near $ h_{c1} $, but become relevant
below the line of fixed points inside the IC-SkX phase in Fig.1.
Eq.\ref{single} has been extended to a finite temperature in \cite{z2}.
The type-I DIO $ u $ leads to the Logarithmic violations of scalings \cite{z2} to the conserved
density $ n = |\psi|^2 $. Then using the relation between the quantum spin and the order parameter in Eq.\ref{k0iyjy}, one can determine the
Logarithmic violations of scalings to various quantum spin correlation functions.

 At the MS point $ \beta=\pi/4 $, $ V=0 $, so the
 effective action Eq.\ref{z2single} is in the same universality class \cite{despite}
  as   the $ z=2 $ zero density SF-Mott transition
 where the interaction $ U $ term is marginally and dangerously irrelevant at the up-critical dimension $ d_u=2 $.
 It  determines the nature of the symmetry breaking ground state and
 also leads to the logarithmic violation of scalings \cite{z2} to all the physical quantities at a finite temperature.

\subsection{ RG and the operator contents away from the MS point }

 When away from the MS point, the $ V $ term in Eq.\ref{z2single}  moves in.
 Simple power counting shows that it is irrelevant near the SF-Mott QPT.
 However, inside the IC-SkX phase, as shown in Eq.\ref{Goldhc1}, it modifies the spectrum of the Goldstone mode
 by an extra linear term, so it plays a crucial role inside the phase.
 We call this new type of DIO as Type-II \cite{type2}, while the known one such as $ U $ as Type-I \cite{z2,rev1,scaling}.
 So there are one Type-I DIO $ u $ and one Type-II DIO $ V $  below the line of stable fixed points
 inside the IC-SkX ( Fig.1 ).


Away from the MS point, one need to add back the type-II dangerously irrelevant operator (DIO) $ V $ in Eq.\ref{z2single},
just by simple power counting, one finds at $ d=2 $:
\begin{equation}
	\partial_l V =- V
\label{singleV}
\end{equation}
  which is clearly irrelevant near $ h_{c1} $, but it  changes the Goldstone mode into the exotic form inside the IC-SkX phase.
  So it becomes marginal below the line of fixed points inside the IC-SkX phase in Fig.1.
  As shown below Eq.\ref{z2single}, the microscopic calculation finds $ V \sim \sin(2 k_0) $ which is indeed exactly marginal along the RG flow in Fig.1
  until hitting the line of fixed points labeled by $ k_0 $ inside the IC-SkX phase.

  So we conclude that away from the MS point, the operator content is:
  one type-I DIO  $ u $  and one type-II  DIO  $ V $.  At the MS point, the operator content is:
  one type-I DIO, no type-II.

 In Fig.1, there are also two trivial line of fixed points at $ h=0 $ and $ h= \infty $  which correspond to $ Z-x $ state and FM state respectively.
 Even it is just the same ground state, the line of fixed points characterized by $ 0 \leq k_0 \leq \pi $  still has clear
 physical meanings: it indicates the minima position of the IC-magnons above the ground state.
 Of course, the Z-x state is an exact ground state, so does not contain the IC-magnons at $ T=0 $.
 However, the lowest excitation is the IC-magnons.
 The FM ground states contains the quantum fluctuations from the IC-magnons even at $ T=0 $.
 So it is justified to use $ k_0 $ to distinguish "different kinds" of Z-x state at $ h=0 $ and "different kinds" of FM at $ h= \infty $ in Fig.1. These could also be one the salient features of In-commensurability  due to SOC.

\section{ Quantum phase transition at the upper critical field $h_{c2}$,
the In-commensurate case }

The SWE in the FM state above $ h_{c2} $ was also performed in \cite{rhh}.
It is the $\alpha_\mathbf{k}$ magnon condensation in Eq.\ref{alphak} which leads to the QPT from the FM state to the IC-SkX at $
 h_{c2} $ in the middle range $\beta_1<\beta<\beta_2$ \cite{middle}:
\begin{equation}
 \alpha_\mathbf{k} =\psi_1\delta_{\mathbf{k},\mathbf{K}_1}
+\psi_2\delta_{\mathbf{k},\mathbf{K}_2}
\label{drop2}
\end{equation}
where $\mathbf{K}_1=(0,k_0), \mathbf{K}_2=(\pi,k_0)$ and $\psi_1, \psi_2$ are the two complex order parameters.

 One must use the Bogoliubov transformation Eq.\ref{bogoliubovTrans} to establish the connection between the transverse quantum spin
 and the two complex order parameters:
\begin{align}
 S_i^+ \propto u[\psi_1+(-1)^{i_x}\psi_2]e^{ik_0i_y}
	+v[\psi_1^\ast-(-1)^{i_x}\psi_2^\ast]e^{-ik_0i_y}
\label{pmk0y}
\end{align}
where $u=u_{\mathbf{K}_1}=u_{\mathbf{K}_2}$
and $v=v_{\mathbf{K}_1}=-v_{\mathbf{K}_2}$.

Because the Z-x state breaks no symmetry of the Hamiltonian, so
one can study how $\psi_1$ and $\psi_2$ transform under the symmetries of the Hamiltonian listed below Eq.\ref{rhh},
especially under  $\mathcal{T}_y $ as $ (\psi_1,\psi_2) \to (e^{ik_0}\psi_1,e^{ik_0}\psi_2) $,
under $ U(1)_{soc} $ as  $ (\psi_1,\psi_2)  \to (\psi_1\cos\phi+i\psi_2\sin\phi,
\psi_2\cos\phi+i\psi_1\sin\phi) $. At the MS point $\beta=\pi/4$, under
$\mathcal{T}\circ\mathcal{M} $ as $(\psi_1,\psi_2) \to(-\psi_1^\ast,-\psi_2^\ast) $.

In fact, as suggested by Eq.\ref{pmk0y}, the physics may become more transparent in the new basis $
	\psi_{\pm}=(\psi_1 \pm \psi_2)/\sqrt{2} $.
   Under the whole family of $\mathcal{T}^n_y, n=1,2,3.......,(\psi_+,\psi_-) \to (e^{i k_0 n }\psi_+,e^{i k_0 n}\psi_-)$.
   When $k_0/\pi$ is an irrational number, $ \theta_0= k_0 n $ becomes a continuous variable leading to a new emergent $ U(1)_{ic} $ symmetry.
   Under $ U(1)_{soc},(\psi_+,\psi_-)\to(e^{i\phi_0}\psi_+,e^{-i\phi_0}\psi_-)$.
   The symmetry analysis implies the effective action:
\begin{align}
	\mathcal{L}_{h_{c2}}
	&=\sum_{\alpha=\pm} (\psi_\alpha^\ast\partial_\tau\psi_\alpha
	+v_x^2|\partial_x\psi_\alpha|^2
	+v_y^2|\partial_y\psi_\alpha|^2	-\mu|\psi_\alpha|^2 )  \nonumber\\
	&+U(|\psi_+|^2+|\psi_-|^2)^2 -A(|\psi_+|^2-|\psi_-|^2)^2 + \cdots \nonumber\\
	&+iV_1(|\psi_+|^2+|\psi_-|^2)
	(\psi_+^\ast\partial_y\psi_++\psi_-^\ast\partial_y\psi_-)  \nonumber\\
	&+iV_2(|\psi_+|^2-|\psi_-|^2)
	(\psi_+\partial_y\psi_+^\ast-\psi_-^\ast\partial_y\psi_-)
\label{pmbasis}
\end{align}
which enjoys a  $ U(1)_{soc} \times U(1)_{ic} $  symmetry when $k_0/\pi$ is an irrational number.
Our microscopic calculation shows that $ \mu=h_{c2}-h, U=h(u^2+v^2)^2+2(1+h) > A=(4+h) > 0 $.
Furthermore, $ V_1, V_2 \propto\sin (2k_0)$, both of which vanish at $ \beta=\pi/4 $ dictated by the MS.

When $\mu=h_{c2}-h <0$, it is in the Z-FM phase with $ \langle \psi_1 \rangle = \langle \psi_2 \rangle=0$.
Expanding the effective action upto the second order in $\psi_\alpha$ leads to 2 degenerate gapped modes
$ \omega_{1,2}=-\mu+v_x^2k_x^2+v_y^2k_y^2 $ which matches the result achieved by SWE in \cite{rhh}.

When $\mu>0$, it is in the IC-SkX phase with $ \langle \psi_1 \rangle= \pm \langle \psi_2 \rangle=\sqrt{\rho_0/2}e^{i\phi_0}$
where $ \rho_0=\sqrt{\mu/2(U-A)}$ ( equivalently $ \langle\psi_-\rangle=0, \langle\psi_+\rangle \neq 0 $ or
$  \langle\psi_-\rangle \neq 0, \langle\psi_+\rangle=0  $ ).
It is easy to see the symmetry breaking pattern is described by the coset:
\begin{equation}
 U(1)_{soc} \times U(1)_{ic}/[U(1)_{soc} \times U(1)_{ic}]_D
\label{coset}
\end{equation}
where the diagonal ( D ) means $ y \to y+n, \phi_0 \to \phi_0 - n k^0_y $
generated by $ \mathcal{T}^{n}_y \times \mathcal{R}( n k^0_y ) $ for any integer $ n $.
The coset dictates only one Goldstone mode ( see Eq.\ref{GoldRoton} ).


Plugging the mean field solution into Eq.\ref{pmk0y},
one obtain the spin-orbital order of the IC-SkX phase below $ h_{c2} $
\begin{equation}
   \langle S_i^+ \rangle=\sqrt{\rho_0/2}[u+v+(-1)^{i_x}(u-v)]e^{(-1)^{i_x}i(k_0i_y+\phi_0)}
\label{orderuv}
\end{equation}
 where $ \langle S_i^z \rangle $ can be fixed by the constraint $|\mathbf{S}_i|^2=S^2$.
It takes the identical form as Eq.\ref{ordercs} after replacing the
Bogliubov transformation matrix elements $ u,v$ by the unitary transformation matrix elements $ c,s $.
It is remarkable that one can extend the unitary transformation matrix elements $ c,s $ in the Z-x phase above $ h_{c1} $ and
the Bogliubov transformation matrix elements $ u, v $ in the FM state below $ h_{c2} $ and reach the same spin-orbital
structure of the IC-SkX phase \cite{sameasEq10}.
Well inside the IC-SkX phase,
one can identify one exotic gapless Goldstone and one exotic gapped roton mode
\begin{eqnarray}
    \omega_{+,\mathbf{k}}
	&=&\sqrt{4\rho_0(U-A)(v_x^2k_x^2+v_y^2k_y^2)}
	-V_{+} \rho_0k_y,  ~~~     \nonumber  \\
    \omega_{-,\mathbf{k}}
	&= & \sqrt{\Delta^2_{-}+8\rho_0A(v_x^2k_x^2+v_y^2k_y^2)}
	-V_{-} \rho_0k_y
\label{GoldRoton}
\end{eqnarray}
  where $ \Delta_{-}= 4\rho_0 A $ is the roton gap, $ V_{\pm}= 4 V_1 \pm 2V_2 $. One can see that
  the Goldstone mode achieved from below $ h_{c2} $ takes the same form as that in Eq.\ref{Goldhc1}
  achieved from above $ h_{c1} $. While the gapped roton mode corresponds to the higher branch $\alpha_\mathbf{k}$
  in Eq.\ref{drop} which is ignored in the effective action Eq.\ref{z2single}.
  This match is a good check on the consistency between  the effective action from $ h_{c2} $ down and that from $ h_{c1} $ up.


In teh following, we study the RG on the two-component QPT near $ h_{c2} $ in Fig.1 and also
list the leading operator contents away from the MS point.
We first derive the RG flow equations at the MS point with $ w \neq 0 $, then
setting $ w=0 $ to study its RG flow, while put the RG flow of $ w \neq 0 $ to next section.

\subsection{ The derivation of the RG flow equations with $ w \neq 0 $. }

The effective action describing  the Magnon BEC in the presence of SOC
near $ h=h_{c2} $ at the Mirror symmetric (MS) point $ \beta= \pi/4 $ in Fig.1 is:
\begin{align}
	&S_{MS} =\int d\tau d^dr \sum_{\alpha=1,2}
	[\psi_\alpha^*(r,\tau)\partial_\tau\psi_\alpha(r,\tau)
	  \nonumber  \\
	&+\frac{\hbar^2}{2m}|\nabla\psi_\alpha(r,\tau)|^2-\mu|\psi_\alpha(r,\tau)|^2]                 \nonumber  \\
    &+ u(|\psi_1(r,\tau)|^4+|\psi_2(r,\tau)|^4)   + v|\psi_1(r,\tau)|^2|\psi_2(r,\tau)|^2    \nonumber  \\
	&+w(\psi_1(r,\tau)\psi_2(r,\tau))^2	+w^*(\psi_1^*(r,\tau)\psi_2^*(r,\tau))^2+ \cdots ]
\end{align}
where $\psi_1$ and $\psi_2$ are two complex scalar fields. From Eq.\ref{pmbasisM}, one can identify $ u=U-A >0, v/2=U+A > 0 $ and $ w=B_2 $.
In fact, $ w $ can be made real and its
the sign  can be changed simply by performing the transformation $ \psi_1 \rightarrow  \psi_1 e^{i \pi/2 } $.
So the sign of $ w $ is irrelevant.
Then the three parameters satisfy $ A > 2 B_2 $, namely $ v/2- u > 4 w $.
After adding back the two type-II dangerously irrelevant operator (DIO) $ V_1 $ and $ V_2 $ in Eq.\ref{pmbasis},
it also describes the QPT near $ h_{c2} $ away from the mirror symmetric point in Fig.1.

\begin{figure}
\includegraphics[width=0.45\textwidth]{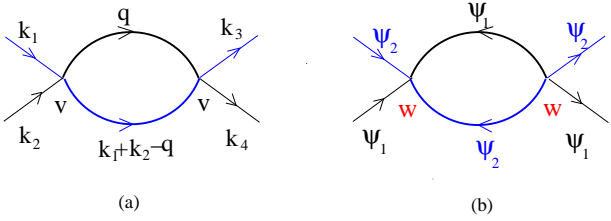}
\caption{ The renormalization of the mutual interaction $ v $ between the two complex order parameters. The same notations as in Fig.\ref{oneloopu}.
The 3-momentum $ k=(\vec{k}, \omega ) $ is indicated in (a) and
The black line and  the green line in (b) means $ \psi_1 $ and $ \psi_2 $ respectively.
(a) due to $ v^2 $ (b) due to $ w^2 $. }
\label{oneloopv}
\end{figure}

\begin{figure}
\includegraphics[width=0.45\textwidth]{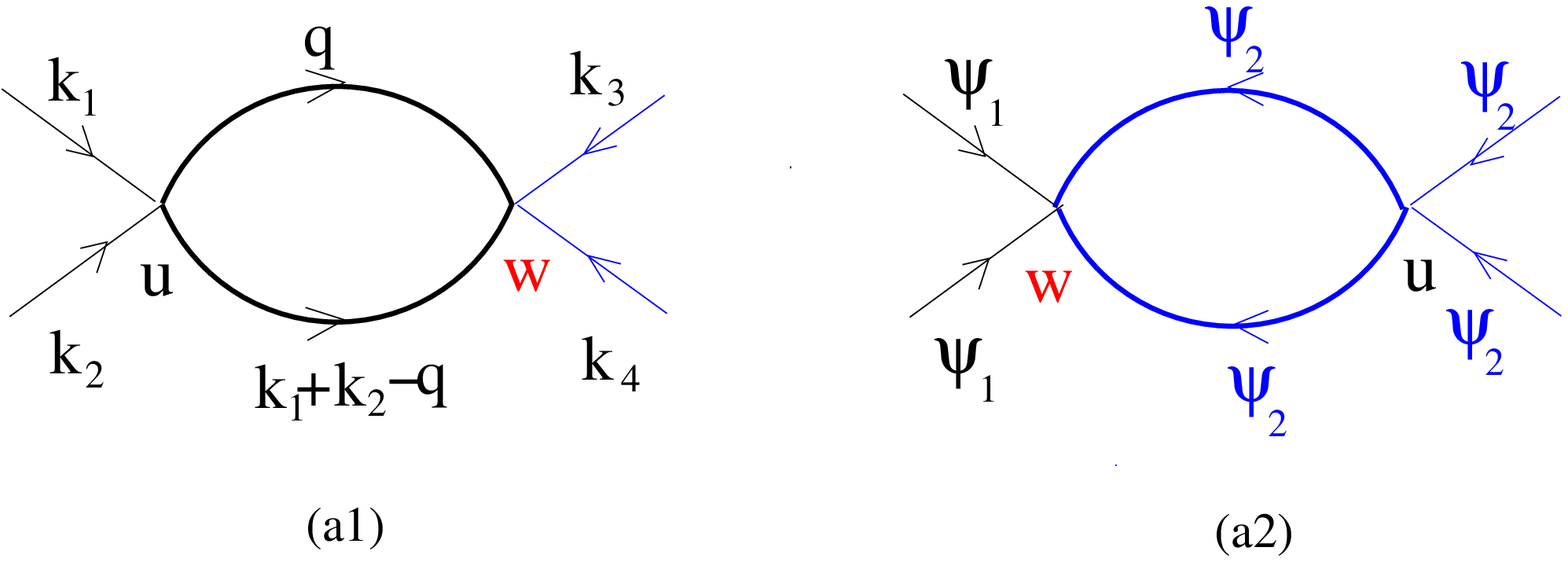}
\includegraphics[width=0.22\textwidth]{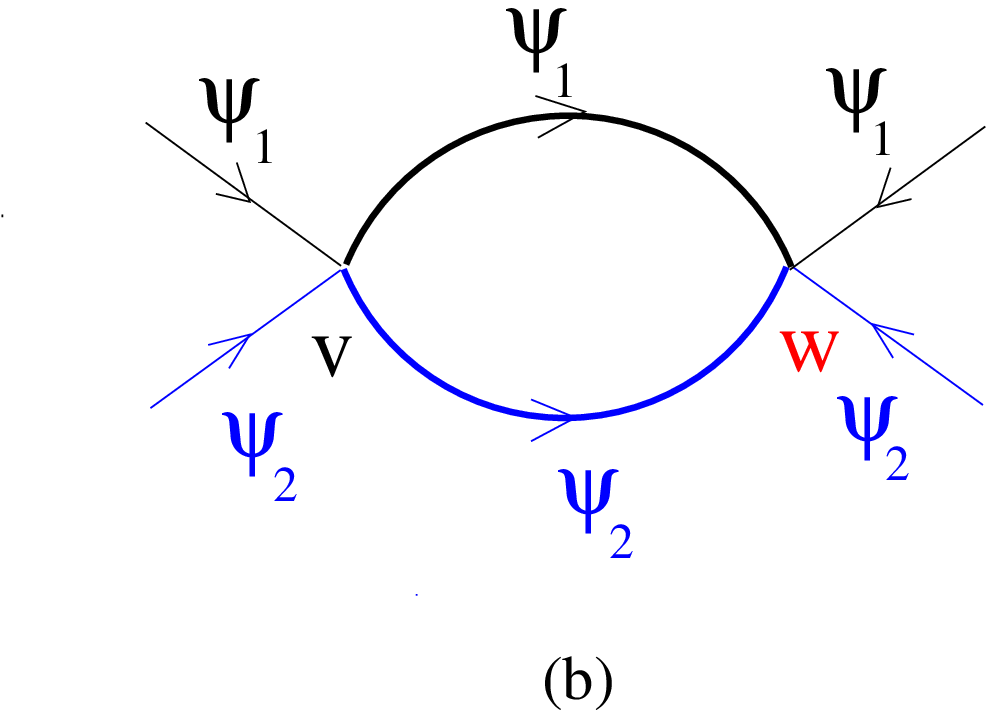}
\caption{  The renormalization of the Umklapp interaction $ w $.
(a1) and (a2) due to $ uw $ (b) due to $ vw $.  }
\label{oneloopw}
\end{figure}

The Wilsonian RG equations upto 1-loop are shown in Fig.\ref{oneloopu}, \ref{oneloopv}, \ref{oneloopw} and collected as:
\begin{eqnarray}
	\partial_l\mu&=&2\mu    \nonumber\\
	\partial_l u &=&\epsilon u- c(u^2+w^2)   \nonumber\\
	\partial_l v &=&\epsilon v- c(\frac{1}{2}v^2+8w^2 )   \nonumber\\
	\partial_l w &=& \epsilon w- 2c(u+v)w
\label{uvw}
\end{eqnarray}
where $\epsilon=2-d$ and $c=K_d\Lambda^{d-2}2m/\hbar$ is identical to that in the single component case Eq.\ref{single}.
One can see that the $ w^2 $ terms renormalize the $ u $ and $ v $, but not itself \cite{not}.
As expected, the RG flow remains the same when $ w \rightarrow -w $.
The application to the MS case will be presented in Sec.IV.

\subsection{ The RG flow pattern and operator contents at $ w = 0 $. }

As said in the main text, the $ w $ term breaks $ U(1)_{ic} \rightarrow Z_4 $, so $ w=0 $ must be a fixed point.
If the umklapp term is absent, $w=0$, then the 1-loop RG equations is simplified to:
\begin{eqnarray}
	\partial_l\mu&= & 2\mu   \nonumber\\
	\partial_l u &= & \epsilon u- cu^2   \nonumber\\
	\partial_l v &= & \epsilon v- cv^2/2
\label{uv}
\end{eqnarray}
The RG flow of $ u $ and $ v $  decouples at the one-loop order.
At $ d=2 $, the RG flow pattern $ 1/u- 2/v= C $
is given in the $ w=0 $ plane of Fig.\ref{fig:RG-PF}a. Both are marginally irrelevant at $ d=2 $
and lead to new logarithmic violations of scalings to all the physical quantities
at a finite $ T $.

In fact, if one rescales $v$ by $1/2$, then $v$ has the same flow equation as $u$.
This is expected, because after setting $ w=0 $, the $ u $ term is the self-interaction, while
the $ v $ term is the only the mutual coupling between the two
complex fields, so $ v=0 $ must be a fixed point also.
Naively, one may think Eq.\ref{pmbasis} has an enlarged $ O(4) $ symmetry at $ A=0 $. This is incorrect due to the $ z=2 $ term in Eq.\ref{pmbasis} which explicitly breaks the $ O(4) $ symmetry.
If it had been $ z=1 $, it would have the enlarged $ O(4) $ symmetry at $ A=0 $. This is just one aspect of the tremendous
differences between $ z=1 $ and $ z=2 $.
In fact we expect that Eq.\ref{uv} is exact to any loop order due to the exact $ U(1)_{soc} \times U(1)_{ic} $ symmetry and $ z=2 $.
Our field theory RG analysis \cite{RGSF} confirm this expectation upto two loops.

In the present case, $ u > 0, v>0 $ satisfying $ v/2 > u $, so both $ u $ and $ v $ are type-I marginally irrelevant near $ h_{c2} $,
but become relevant above the line of fixed points inside the IC-SkX phase in Fig. 1.
It is important to extend Eq.\ref{uv} to a finite temperature. Then the two type-I DIOs $ u $ and $ v $ lead to
Logarithmic violations of scalings to the two conserved quantities $ n_1 = |\psi_1|^2 $ and $ n_2 = |\psi_2|^2 $
( or equivalently  $ |\psi_1|^2  + |\psi_2|^2 $ or $ |\psi_1|^2 - |\psi_2|^2 $ ). Then using Eq.\ref{pmk0y}, one can determine the
Logarithmic violations of scalings to various quantum spin quantities.

The $ v > 0 $ means the repulsive interaction between the two species.
Imagine the inter-specie interaction becomes negative $ v < 0 $, Eq.\ref{uv} shows that it flows to $ - \infty $
which indicates the formation of the tightly formed bosonic Cooper pair between $ \psi_1 $ and $ \psi_2 $.

After adding back the two type-II dangerously irrelevant operator (DIO) $ V_1 $ and $ V_2 $ in Eq.\ref{pmbasis},
just by simple power counting, one finds at $ d=2 $:
\begin{align}
	\partial_l V_1 &=- V_1,   \nonumber  \\
    	\partial_l V_2 &=- V_2
\label{twoV}
\end{align}
  which are clearly irrelevant near $ h_{c2} $, but  change the gapless Goldstone mode and the gapped
  roton into the exotic form inside the IC-SkX phase.
  So $ V_1, V_2 $ become marginal above the line of fixed points inside the IC-SkX phase in Fig. M1.
  As shown below Eq.\ref{pmbasis}, the microscopic calculations find $ V_1, V_2 \sim \sin(2 k_0) $
  which are indeed exactly marginal along the RG flow in Fig.1
  until hitting the line of fixed points labeled by $ k_0 $ inside the IC-SkX phase.

We conclude that away from the MS point near $ h_{c2} $, the operator content is:
two type-I DIOs $ ( u, v ) $ in Eq.\ref{uv} and two type-II DIOs $ ( V_1, V_2 ) $  in Eq.\ref{twoV} above the line of stable fixed points
inside the IC-SkX ( Fig.1 ).

\section{ Quantum phase transition at the upper critical field $h_{c2}$,
the Commensurate  case and MS point }

However, if $k_0/\pi=p/q$ with $p$ and $q$ are two co-prime positive integers,
then $(\mathcal{T}_y)^{2q}=1$ and the effective action Eq.\ref{pmbasis} should include an extra Umklapp term:
\begin{align}
	\mathcal{L}_\text{Um}
	&=B_q(\psi_+\psi_-)^{q}+ c.c.   \nonumber  \\
    &+ iC_q(\psi_+\psi_-)^{q-1}(\psi_+\partial_y\psi_-)+c.c.+ \cdots
\label{umterm}
\end{align}
where $B_q,C_q$  maybe complex for $\beta\neq\pi/4$  and $\cdots$ means high order terms with power $2nq$ ($n>1$).
It breaks explicitly only the $ U(1)_{ic} $ down to $ Z_{2q} $, but not the $ U(1)_{soc} $ symmetry.
In the regime $0\leq k_0\leq\pi/2$ in Fig.1, $q\geq 2$, so $\mathcal{L}_\text{Um}$  becomes higher order
when $\beta<\pi/4$ with $ q>2 $. It becomes highly irrelevant,
so can be dropped. All the above results in the IC-case also apply to $ q>2 $ C- case.

At the MS point $ \beta=\pi/4 $,  $k_0=\pi/2$ with $q=2$,
then $\mathcal{L}_\text{Um}$ is quartic order in $\psi_{1,2}$ .
So one must consider this $ B_2 $ term at $ \beta=\pi/4 $ where the MS dictates $ C_2=0 $ and also the absence of the
two type-II  DIOs $ V_1,V_2 $.
\begin{align}
	\mathcal{L}_{MS}
	 & =\sum_{\alpha=+,-}
	(\psi_\alpha^\ast\partial_\tau\psi_\alpha
	+v_x^2|\partial_x\psi_\alpha|^2
	+v_y^2|\partial_y\psi_\alpha|^2)-\mu|\psi_\alpha|^2 )    \nonumber  \\
	&+U(|\psi_+|^2+|\psi_-|^2)^2   -A(|\psi_+|^2-|\psi_-|^2)^2    \nonumber  \\
    & + B_2 (\psi_+\psi_-)^{2}+c.c. +\cdots
\label{pmbasisM}
\end{align}

 The microscopic calculations in \cite{rhh} show $ A > 2 |B_2| $, so the $ B_2 $ term does not change the mean field state.
 Following the same procedures as those in the IC- case, one can extract the excitations as:
\begin{align}
    \omega_{+,\mathbf{k}}
	&=\sqrt{4\rho_0(U-A)(v_x^2k_x^2+v_y^2k_y^2)},  \nonumber   \\
    \omega_{-,\mathbf{k}}
	&=\sqrt{16\rho_0^2(A^2-4B_2^2)+8\rho_0A(v_x^2k_x^2+v_y^2k_y^2)}
\end{align}
 which recover to the conventional form.
 It also indicate the Umklapp term at $ \beta=\pi/4 $ does not affect the form of the Goldstone mode, but decrease the roton gap.
 This is expected, because this $ B_2 $ term breaks only the $ U(1)_{ic} $ down to $ Z_4 $, but not the  $ U(1)_{soc} $ symmetry.


Now we study the QPT near $ h=h_{c2} $. The RG flow equations at one -loop were already derived in Sec.III-A  Eq.\ref{uvw}
where $ w=B_2 $. $ w=0 $ must be a fixed point.
This is expected, because the $ w $ term breaks $ U(1)_{ic} \rightarrow Z_4 $.
Setting $ w=0 $  recovers Eq.\ref{uv}.
As to be shown below,
in the physically accessible parameter regime $ v/2-u > 4 w >0 $, RG flows to the Gaussian fixed point
$ (u^{*}, v^{*}, w^{*} )=(0,0,0) $. So $ ( u, v, w ) $ all lead to new Logarithmic violations of scalings to all the physical quantities
at a finite $ T $.
So there are three Type-I DIOs, but no Type-II at the MS point near $ h_{c2} $ ( Fig.1 ).

Now we study the RG on the two-component QPT near $ h_{c2} $ in Fig.1 and also work out the elading  operator contents at the MS point.
At the MS point, one can turn on the $ w $ term which breaks $ U(1)_{ic} \rightarrow Z_4 $.
At the upper critical dimension $ d_u=2 $, $\epsilon=0$,
rescale $u,v,w$ by the positive number $c$ in Eq.\ref{uvw},
i.e. $u\to cu$,$v\to cv$,$w\to cw$,
the RG flow equations become
\begin{eqnarray}
	\partial_l u &= &- (u^2+w^2)  \nonumber\\
	\partial_l v &= & - (v^2+16w^2)/2   \nonumber\\
	\partial_l w &= &- 2(u+v)w
\label{uvw00}
\end{eqnarray}
 which gives a 3d flow in $ (u, v, w ) $ space.

For the RG flow Eq.\ref{uvw} with the 3 couplings $ u, v $ and $ w $, it would be nice to find an integral motion.
Unfortunately, in contrast to the case of the three coupling of fermions near a Fermi surface studied  in Ref.\cite{highTc},
we are not able to find any  integral motion of the RG flow of Eq.\ref{uvw}.
In the following, we find its RG flow pattern first analytically for a small initial value of $ w_0 $, then numerically for
any initial values of $ (u_0, v_0, w_0 ) $. Finally, we show that the physically accessible values of  $ (u, v, w ) $, it flows to
the Gaussian fixed point $(u_*,v_*,w_*)=(0,0,0)$.

\subsection{ Analytic perturbation analysis for a small initial $ w_0 $. }

Taking any general initial condition $(u_0,v_0,w_0)$ with $w_0=0$,
Eq.\eqref{uv} can be solved analytically:
\begin{align}
	u(l)=\frac{u_0}{1+u_0 l},\quad
	v(l)=\frac{2v_0}{2+v_0 l},\quad
	w(l)=0
\end{align}

\begin{figure}[tbhp]
	\centering
	\includegraphics[width=0.8\linewidth]{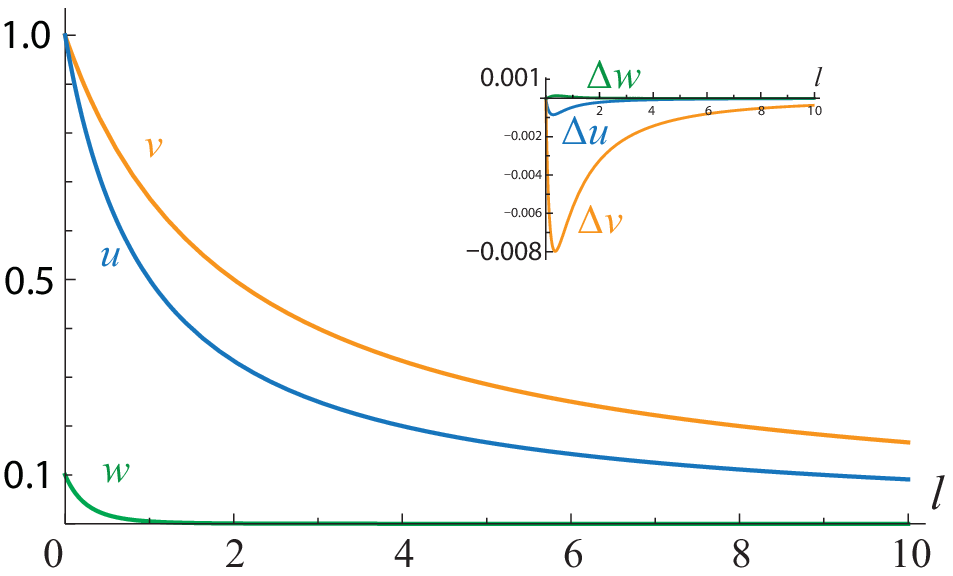}\qquad
	\includegraphics[width=0.8\linewidth]{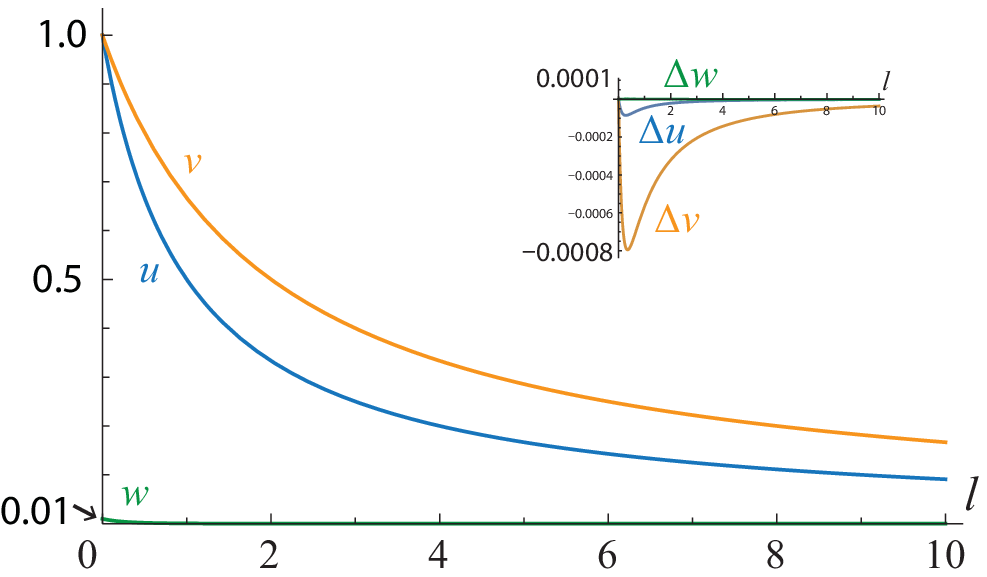}
		\caption{ The solution for RG equation with big initial $u_0,v_0$, but small initial value $w_0$.
		The insert is the differences between numerically solution and analytical solution
		Eq.\eqref{eq:u1v1},\eqref{eq:wl}. The top one is $(u_0,v_0,w_0)=(1,1,0.1)$,
		and the bottom one is $(u_0,v_0,w_0)=(1,1,0.01)$. One can see the differences
		between numerically solution and analytical solution decrease with $w_0\to 0$.}
	\label{fig:rg-12}
\end{figure}

For $|w_0|\ll u_0, v_0$, assume the solution takes the form
\begin{align}
	u(l)=\frac{u_0}{1+u_0 l}+u_1(l),\quad
	v(l)=\frac{2v_0}{2+v_0 l}+v_1(l),\quad
	w(l)=w(l)
\label{eq:u1v1}
\end{align}
where $u_1(l),v_1(l),w(l)\to 0$ as $w_0\to0$.
Ignoring the higher orders $[u_1(l)]^2,[v_1(l)]^2$ terms,
we reach a new set of differential equations
\begin{align}
	\partial_l u_1&=-\frac{2u_0}{1+u_0 l}u_1 \nonumber\\
	\partial_l v_1&=-\frac{2v_0}{2+v_0 l}v_1 \nonumber\\
	\partial_l w&=-2(\frac{u_0}{1+u_0l}+\frac{2v_0}{2+v_0l}+u_1+v_1)w
\end{align}
with initial condition $u_1(0)=0,v_1(0)=0,w(0)=w_0$.
The solution is $u_1=0$, $v_1=0$, and
\begin{align}
	w(l)=\frac{16w_0}{(1+u_0l)^2(2+v_0 l)^4}
	\sim \frac{ 16 w_0}{u^2_0 v^4_0 } \frac{1}{l^4}
\label{eq:wl}
\end{align}
where the last $ \sim $ shows the asymptotic behaviour at a large $ l $.
It is clear that with a small $w_0$ and any  positive $u_0$,$v_0$,
the RG  flows to the Gaussian fixed point.
In Fig.\ref{fig:rg-12}, we plot the differences between the analytical solution and numerical solution,
which agree very well in the small $w_0$ limit.

\subsection{ Numerical solution of the RG flow for any initial values of $ (u_0, v_0, w_0 ) $ }

\begin{figure}[tbhp]
	\centering
	\includegraphics[width=0.8\linewidth]{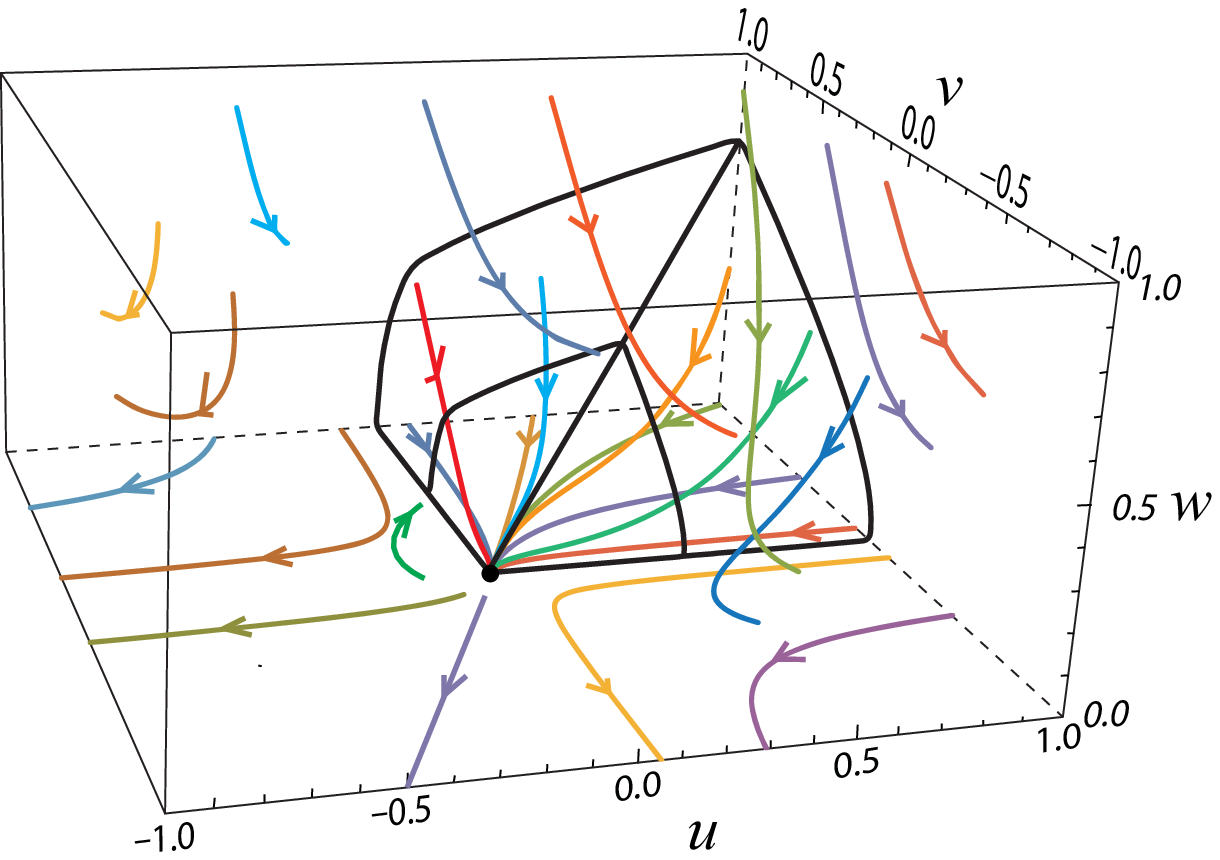}\qquad
	\includegraphics[width=0.8\linewidth]{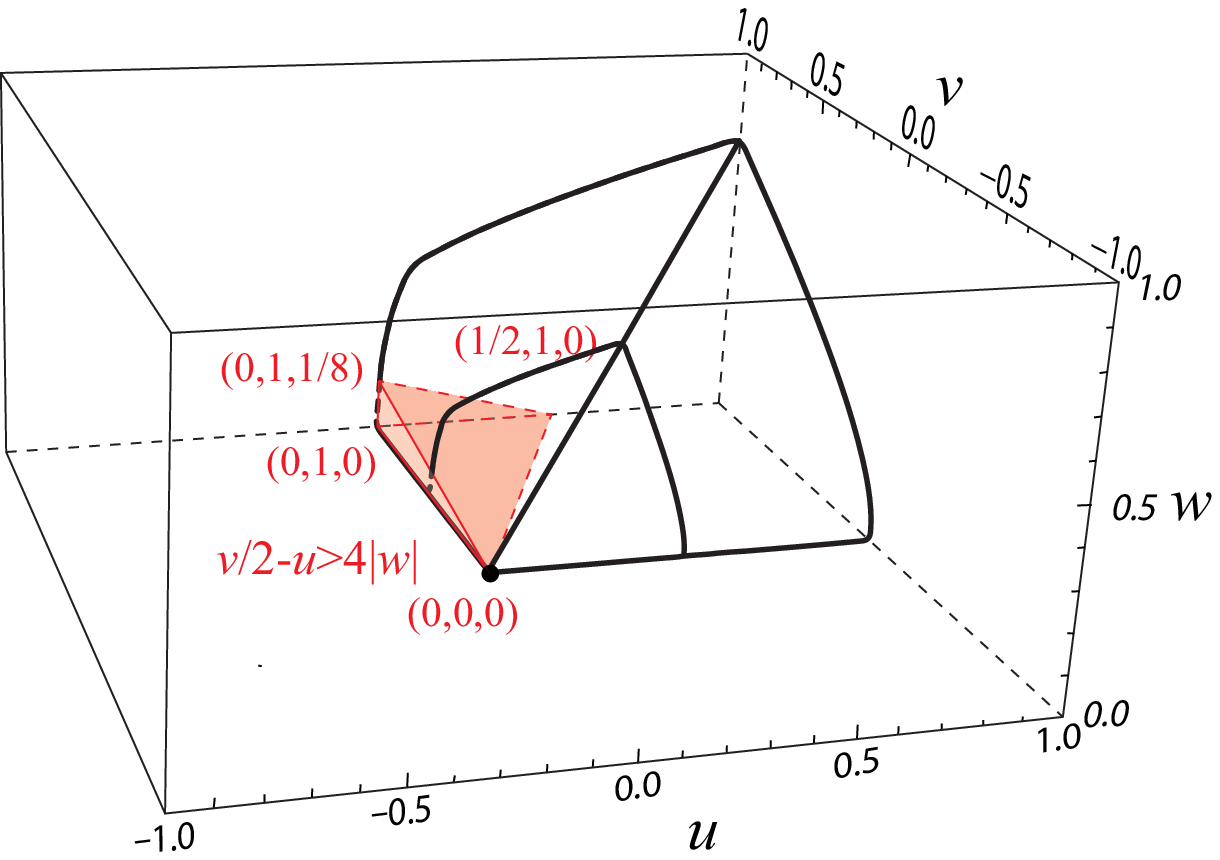}
		\caption{Top (a) The RG flow pattern in the $(u,v,w)$ space. The $ w=0 $ plane recovers the RG flow $ 1/u-2/v= C $ in Eq.\ref{uv}.
        The $ C=0 $ gives the straight line flow $ 1/u=2/v $.
        The black thick lines delineate the regime where the RG flows to the Gaussian fixed point.
		Bottom (b)  The physically accessible  regime (thin red lines) falls inside the stable regime (thick black lines).}
	\label{fig:RG-PF}
\end{figure}

To go beyond the small perturbation presented above,
we numerically solve the differential equations Eq.\ref{uvw00} for any general initial values of $ (u_0, v_0, w_0 ) $
and highlight the parameter regime which flows to the Gaussian fixed point in Fig.\ref{fig:RG-PF}a.
Due to the $w\to -w$ symmetry, we only need plot the flows  with $w>0$,
and then there is a large stable regime of the Gaussian fixed point inside the first Octant.

Then we plot the initial condition wedge $ v/2-u > 4|w|$ and find that it falls within the stability regime in Fig.\ref{fig:RG-PF}b.
This shows that in the physically allowed parameter regime, the systems flows to the Gaussian fixed point $(u_*,v_*,w_*)=(0,0,0)$.
All the three operators $ (u,v,w) $ are type-I dangerously irrelevant which are marginally irrelevant at the MS point in the line of fixed points
along $ h_{c2} $, but relevant above the line of fixed points inside the IC-SkX phase (Fig.1 ).

It is important to extend Eq.\ref{uvw00} to a finite temperature. Then the three type-I DIOs $ (u, v, w ) $
leads to Logarithmic violations of scalings to one conserved quantity  $ |\psi_1|^2  + |\psi_2|^2 $
and also to $ |\psi_1|^2 - |\psi_2|^2 $ which is not conserved anymore due to the $ w $ term breaks
explicitly $ U(1)_{ic} \rightarrow  Z_4 $. Then using Eq.\ref{pmk0y}, one can determine the
Logarithmic violations of scalings to various quantum spin quantities.

We conclude that at the MS point near $ h_{c2} $, the operator content is:  three type-I DIOs $ ( u, v, w ) $, none type-II.

\section{  Discussions and Conclusions  }

In this work,
by  constructing effective actions from general symmetry principle, contrasting with the previous
microscopic calculations well inside a phase  and performing RG analysis near the QCPs,
we study the BEC of magnons in the presence of SOC which displays much more richer and novel phenomena than those without SOC.

From symmetry analysis, plus some inputs from the  microscopic SWE calculations achieved in \cite{rh,rhh},
we constructed various effective actions to describe the two transitions driven by the BEC of magnons with the dynamic exponents $ z=2 $
at the low and high critical fields. There are one and two type-II dangerously irrelevant operators which lead to exotic
excitations inside the IC-SkX phase.
(1) The C-IC transition from the  Z-x to the IC-SkX at $ h_{c1} $: it has
   one complex order parameter  and one Type-II dangerously irrelevant operator.
(2) The C-IC transition from  the Z-FM to the IC-SkX at $ h_{c2} $ in the middle of SOC $ \beta_1 < \beta < \beta_2 $:
   It has two complex order parameters with the dynamic exponents $ z=2 $ and two Type-II dangerously irrelevant operators.

The RG flows, operator contents, novel line of fixed points labeled by the IC-momenta $ 0 < k_0 < \pi $ in Fig.1 brings out
deep and profound connections between $ 2+1 $ dimensional
non-relativistic quantum field theories (QFTs) with $ z=2 $ and the nature of the quantum magnets with SOC in a Zeeman field.
It remains interesting to construct a F-theorem \cite{EE} to characterize the RG flow from the
UV unstable line of fixed points at $ h_{c1} $ or $ h_{c2} $
to the stable IR line of fixed points labeled by $ k_0 $ inside the IC-SkX phase  for such novel QFTs.

As argued in \cite{rh}, Eq.1 could be easily realized in recent cold atom experiments
\cite{expk40,expk40zeeman,clock,2dsocbec,ben} to
generate 2D Rashba SOC for cold atoms on optical lattices in a Zeeman field.
The spin-orbital structure of the IC-SkX and its exotic excitations in Fig.1
can be directly detected by all kinds of Bragg spectroscopy \cite{lightatom1,braggbog}.
The IC-SkX phase was realized in some materials with a strong Dzyaloshinskii-Moriya (DM) interaction.
Indeed, a 2D skyrmion lattice has been observed between $ h_{c1}=50$ mT  and $ h_{c2}=70$ mT
in some chiral magnets \cite{sky4} MnSi or a thin film of Fe$_{0.5}$Co$_{0.5}$Si \cite{sky4}.
Fig.1 also suggests the "exotic" intermediate phase observed between $ h_{c1}=7$ T  and $ h_{c2}=9$ T
in so called 4d Kitaev material $ \alpha-Ru Cl_3 $ \cite{halfinteger,unquantized}
could be nothing but just the IC-SkX phase discovered here instead of a quantum spin liquid phase.
It may also shed some lights on the magnon condensation in some dimerized antiferromagnets
with SOC \cite{dimer}.

In conventional quantum magnets, as discussed in the appendix B, the $ U(1)_s $ symmetry is only an approximation which can be explicitly broken by
many interactions \cite{frusrev,rev1,rev2}. Here, the existence of $ U(1)_{soc} $ plays crucial role in these phenomena which
only holds along the $ (\alpha=\pi/2, \beta) $ SOC line and the longitudinal Zeeman field.
there could also be many ways to break the $ U(1)_{soc} $ symmetry explicitly.
One way is to apply transverse field.
Another is to look at a generic  $ (\alpha,\beta) $, or one can apply both at the same time.
In \cite{devil}, we showed that the Z-x state remains stable in a large SOC parameter regime near $  \alpha=\pi/2 $,
just changes from the exact to the classical ground state.
In fact, it is the most robust quantum phase in the whole global phase diagram in the generic $ (\alpha,\beta) $.
It would be interesting to look at how Fig.1, especially the intermediate IC-SkX phase changes
when the $ U(1)_{soc} $ symmetry was explicitly broken.
It is also important to extend it to the honeycomb lattice where there are 3 SOC order parameters $ ( \alpha,\beta,\gamma ) $.
It was found that $ U(1)_{soc} $ also exists along the anisotropic line $  ( \alpha=\pi/2,\beta,\gamma=0 ) $.
Then the results to be achieved in \cite{unpublish} may be directly relevant to the current trends
to search for QSL driven by a Zeeman field in 4d or 5d Kitaev materials.

{\bf Acknowledgement }

We thank Prof. Gang Tian and Prof. Congjun Wu  for the hospitality during their visit at the West Lake university.
J. Ye thank Prof. Gang Tian  for the hospitality during his visit at the Great Bay university.

\appendix

\section{Spin-wave expansion to order $1/S $: Unitary transformation and Bogoliubov transformation  }

  We first review some results from spin-wave expansion (SWE) performed in \cite{rhh}.
  Especially, we stress the unitary transformation in the Z-x state below $ h_{c1} $ and the
  Bogoliubov transformation in the  FM state above $ h_{c2} $
  which are crucial to derive the relations between the quantum spin and the order parameters inside
  the IC-SkX phase. As presented in Sec.2 and 3,4, the former is from bottom-up and the latter is from top-down.

\subsection{ Unitary transformation in the Z-x state in low field}

The spin $ S=N/2 $ Rotated Ferromagnetic Heisenberg model
at a generic SOC parameters $ ( \alpha, \beta) $ in a Zeeman field $ \vec{H} $ along any direction is \cite{rh}:
\begin{align}
	\mathcal{H}_{RH} & =  -J\sum_i
	[\mathbf{S}_i R(\hat{x},2\alpha)\mathbf{S}_{i+\hat{x}}
	+\mathbf{S}_i R(\hat{y},2\beta)\mathbf{S}_{i+\hat{y}}]
   \nonumber   \\
     & -  \vec{H} \cdot \sum_i \vec{S}
\label{rhgeneral}
\end{align}
where  the $ R(\hat{x}, 2 \alpha),  R(\hat{y}, 2 \beta)$ are
two $ SO(3) $ rotation matrices around  the $ \hat{x}, \hat{y} $  spin axis by angle $ 2 \alpha, 2 \beta $
putting along the two  bonds  $ x,y $ respectively, $H$ is the  Zeeman field
which could be induced by the Raman laser in the cold atom set-ups.

Following \cite{rhh}, we focus on studying the phenomena along the line $ (\alpha=\pi/2, 0<\beta<\pi/2 ) $
and in the Zeeman field along the longitudinal $ y $ direction.
After rotating spin $Y$  axis to $Z$ axis by the global rotation  $R(\hat{x},\pi/2)$,
(or equivalently, one can just put $ \beta \sigma_z $  along the $ y $ bonds in the square lattice ),
the Hamiltonian Eq..\ref{rhgeneral} along the line $ ( \alpha=\pi/2, 0<\beta<\pi/2) $
in the $ H $ field along $ y $ direction can be written as Eq. M1.

 In a weak magnetic field $ h < h_{c1} $, the Z-x state is an exact ground-state with the classical spin configuration:
\begin{align}
	\mathbf{S}_i=S(0,0,(-1)^{i_x})
\end{align}
  which breaks the translational symmetry along the x-bond to two sites per unit cell ( so named as Z-x phase ).

Performing the Holstein-Primakoff (HP) transformation
for the sublattice A \cite{rhh}
\begin{align}
	S_i^+ & =\sqrt{2S}\Big(1-\frac{1}{2}\frac{n_i}{2S}+\cdots\Big)a_i,\quad           \nonumber  \\
	S_i^- & =\sqrt{2S}a_i^\dagger\Big(1-\frac{1}{2}\frac{n_i}{2S}+\cdots\Big),\quad    \nonumber  \\
	S_i^z & =S-a_i^\dagger a_i,\quad i\in A;
\end{align}
  and for the sublattice B
\begin{align}
    S_j^+ & =\sqrt{2S}b_j^\dagger\Big(1-\frac{1}{2}\frac{n_j}{2S}+\cdots\Big),\quad      \nonumber  \\
	S_j^- & =\sqrt{2S}\Big(1-\frac{1}{2}\frac{n_j}{2S}+\cdots\Big)b_j,\quad              \nonumber  \\
	S_j^z & =-S+b_j^\dagger b_j,\quad j\in B.
\end{align}

In momentum space, the Hamiltonian takes the form
\begin{align}
	\mathcal{H}&=-2NJs^2
	+H\sum_k(a_k^\dagger a_k-b_k^\dagger b_k)
	+4JS\sum_k(a_k^\dagger a_k+b_k^\dagger b_k)  \nonumber  \\
	&-2JS\sum_k[\cos k_x a_k^\dagger b_k+\cos k_x b_k^\dagger a_k   \nonumber  \\
	& +\cos(k_y-2\beta)a_k^\dagger a_k      	+\cos(k_y+2\beta)b_k^\dagger b_k]
\end{align}
 By performing a unitary transformation
\begin{align}
	a_\mathbf{k}=s_\mathbf{k}\alpha_\mathbf{k}
	+c_\mathbf{k}\beta_\mathbf{k},\quad
	b_\mathbf{k}=s_\mathbf{k}\beta_\mathbf{k}
	-c_\mathbf{k}\alpha_\mathbf{k},
\label{unitaryTrans}
\end{align}
where $s_\mathbf{k}=\sin(\theta_{k,h}/2)$, $c_\mathbf{k}=\cos(\theta_{k,h}/2)$ satisfying  $ c^2 + s^2 = 1 $,
and $\tan\theta_{k,h}=\cos k_x/(\sin2\beta\sin k_y-h)$, the Hamiltonian can be put in the diagonal form:
\begin{align}
	\mathcal{H}=-2NJS^2
	+4JS\sum_k[\omega_{+}(k)\alpha_k^\dagger\alpha_k+\omega_{-}(k)\beta_k^\dagger\beta_k]
\label{betak}
\end{align}
where $ k $ is in the reduced BZ and the excitation spectrum is:
\begin{align}
	\omega_{\pm}(k) & =1-\frac{1}{2}\cos2\beta\cos k_y
                          \nonumber  \\
	  & \pm \frac{1}{2}\sqrt{\cos^2 k_x+(\sin 2\beta\sin k_y-h)^2}.
\end{align}

  The unitary transformation matrix elements $s_\mathbf{k}=\sin(\theta_{k,h}/2)$ and $c_\mathbf{k}=\cos(\theta_{k,h}/2)$
  are useful to establish the connections between the transverse quantum spin and the order parameter near $ h_{c1} $ in
  Eq.\ref{k0iyjy}.

\begin{figure}
\includegraphics[width=0.32\textwidth]{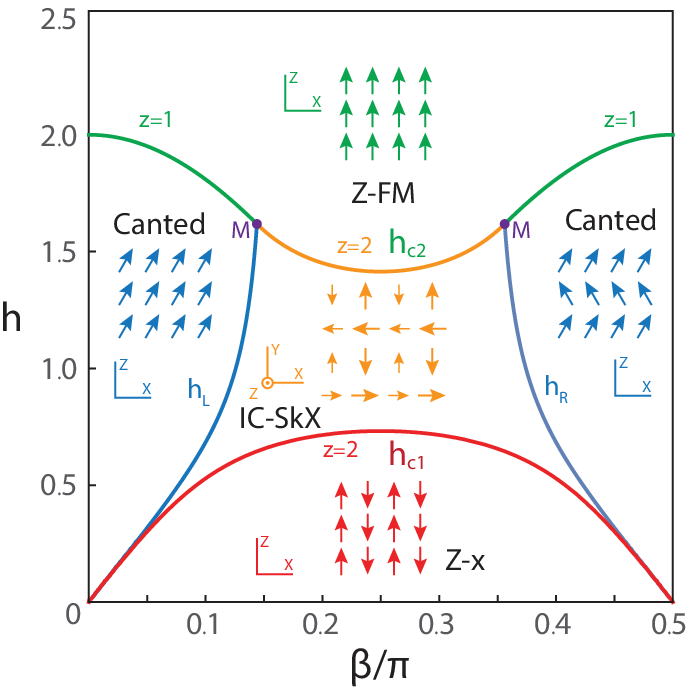}
\hspace{1cm}
\includegraphics[width=0.4\textwidth]{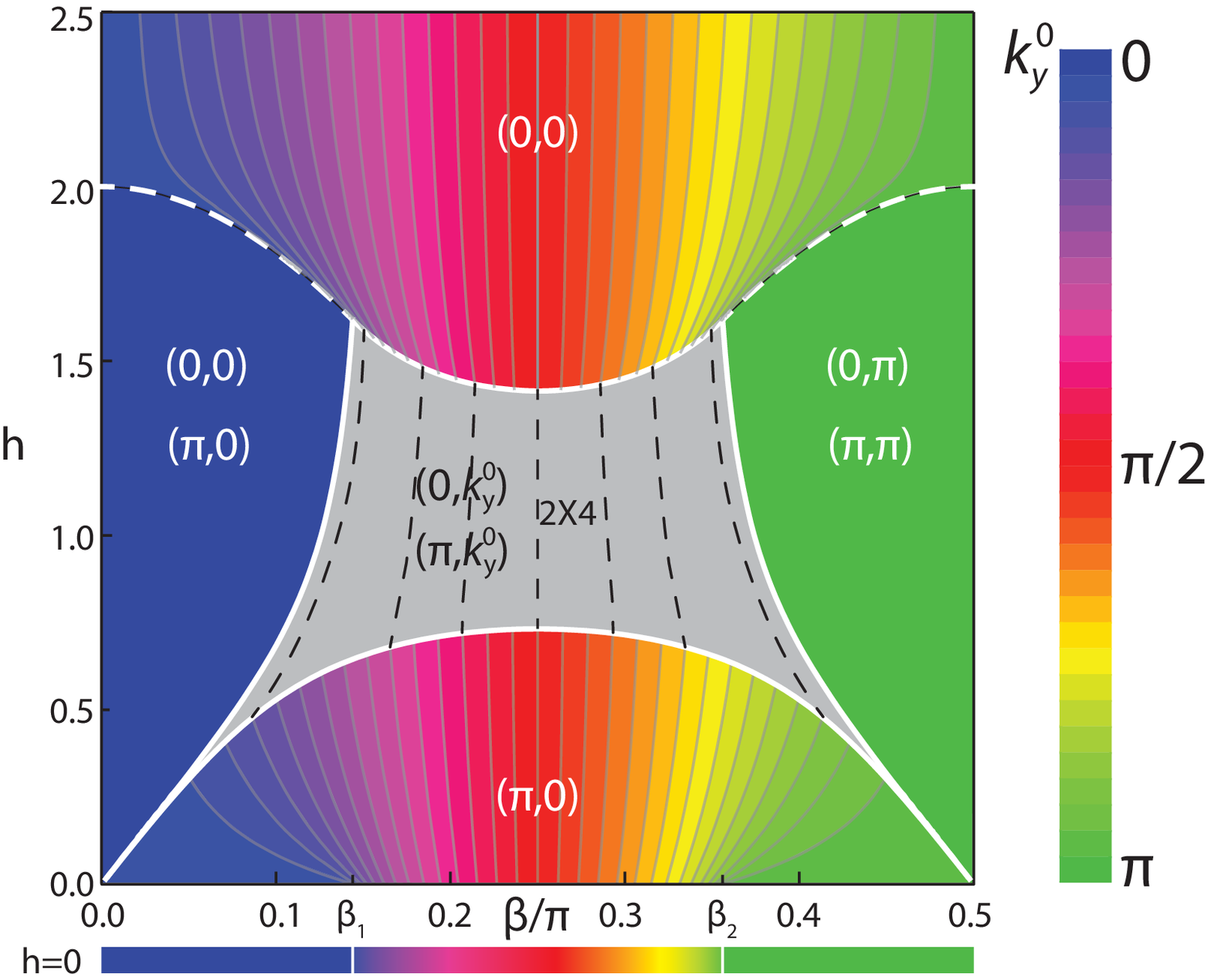}
\caption{ Top (a) and bottom (b) are achieved in \cite{rhh}
(a) Quantum phases and quantum transitions of RFHM in a longitudinal Zeeman field Eq.1 achieved by the
 microscopic SWE in  \cite{rhh}.
The main text focus on constructing various effective actions and then performing RG analysis on them to study
the QPT  from the Z-x to the IC-SkX at $ h_{c1} $ and from the Z-FM to the IC-SkX at $ h_{c2} $ at and away from
the Mirror Symmetric (MS) point at $ \beta=\pi/4 $.
There are one and two type-II dangerous irrelevant operators respectively which disappear at the MS point.
(b)  The orbital ordering  wavevectors of  the two collinear, two coplanar and the non-coplanar phases.
The constant contour plot of the minima $ (0, k_y^0 ) $  of the C-IC magnons in the Z-x state at $ h < h_{c1} $ and  Z-FM state
at $ h>h_{c2} $, connected by the orbital ordering wavevectors ( dashed line ) inside the IC-SkX. }
\label{AFMh}
\end{figure}

\subsection{ Bogoliubov transformation in the FM in the high field}
In a strong magnetic field, the Z-FM state is the classical ground-state  with the classical spin configuration:
\begin{align}
	\mathbf{S}_i=S(0,0,1)
\end{align}
Performing the standard HP transformation
\begin{align}
	S_i^+ & =\sqrt{2S}\Big(1-\frac{1}{2}\frac{n_i}{2S}+\cdots\Big)a_i,\quad                  \nonumber  \\
	S_i^- & =\sqrt{2S}a_i^\dagger\Big(1-\frac{1}{2}\frac{n_i}{2S}+\cdots\Big),\quad          \nonumber  \\
	S_i^z & =S-a_i^\dagger a_i.
\end{align}
In momentum space, the Hamiltonian takes the form
\begin{align}
	\mathcal{H} & =-NHS+H\sum_{k}a_k^\dagger a_k
	-JS\sum_k[2\cos (k_y-2\beta)a_k^\dagger a_{k}       \nonumber  \\
	 & +\cos k_x(a_ka_{-k}+a_k^\dagger a_{-k}^\dagger)]
\end{align}
 By introducing the Bogoliubov transformation as
\begin{align}
	a_\mathbf{k}=u_\mathbf{k}\alpha_\mathbf{k}
	+v_\mathbf{k}\alpha_{-\mathbf{k}}^\dagger,\quad
	a_{-\mathbf{k}}^\dagger=v_\mathbf{k}\alpha_{\mathbf{k}}
	+u_\mathbf{k}\alpha_{-\mathbf{k}}^\dagger.
\label{bogoliubovTrans}
\end{align}
where $u_\mathbf{k}=\cosh\eta_k$, $v_\mathbf{k}=\sinh\eta_k$  satisfying  $ u^2 - v^2 = 1 $ and
$\tanh2\eta_k=\cos k_x/(h-\cos2\beta\cos k_y)$,
the Hamiltonian takes the diagonal form
\begin{eqnarray}
	\mathcal{H}&=&-NH(S+\frac{1}{2})+JS\sum_k
	[\omega(k)\alpha_k^\dagger \alpha_k
	+\omega(-k)\alpha_{-k}\alpha_{-k}^\dagger]   \nonumber\\
	&=&-NH(S+\frac{1}{2})+JS\sum_k \omega_k
	+2JS\sum_k \omega_k\alpha_k^\dagger \alpha_k
\label{alphak}
\end{eqnarray}
where $ k $ is in the BZ and the spin wave dispersion is
\begin{equation}
	\omega_k=\sqrt{(h-\cos2\beta\cos k_y)^2-\cos^2 k_x}-\sin2\beta\sin k_y
\label{eq:Ek}
\end{equation}

 The Bogoliubov transformation matrix elements $u_\mathbf{k}$ and $v_\mathbf{k} $
 are useful to establish the connections between the transverse quantum spin and the order parameter near $ h_{c2} $ in Eq.\ref{pmk0y}.


\vspace{0.3cm}

\begin{figure}
\includegraphics[width=0.4\textwidth]{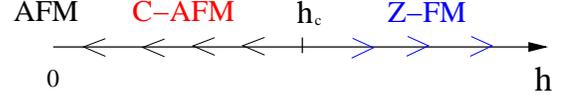}
\caption{ Quantum phase transitions of a AFM in a Zeeman field. C-AFM means the canted co-planar AFM state, Z-FM means the FM along the Zeeman field direction. There is only one critical field $ h_c $, no intermediate phase. The arrow indicate the RG flow, there is only one unstable (QPT)
fixed point at $ h_{c} $. It is constructive to compare with the MS line with $ \beta=\pi/4 $ in (a) and RG flow in Fig. M1.  }
\label{AFMh0}
\end{figure}

\section{ The BEC of magnons of an AFM in a uniform field: the $ U(1)_s $ case }

For an AFM in a uniform field which breaks the spin $ SU_s(2) $ to the spin $ U(1)_s $.
A high $h > h_c $ leads to a fully polarized state,
Z-FM state, which is not only the ground state but also an exact eigenstate.
A Simple spin-wave calculation shows $\omega\sim \Delta+ v^2 k^2$ near $(\pi,\pi)$ which is nothing but a gapped FM mode,
the order parameter is simply  a complex field $\psi$.
Because the Z-FM is the exact ground state which does not break the translational symmetry,
neither Bogoliubov transformation nor unitary transformation is needed,
thus the relation between the  quantum spin in a square lattice and the order parameter in a continuum is simply
\begin{equation}
  \langle S_i^+\rangle= (-1)^{i_x+i_y}\psi
\label{u1s}
\end{equation}
   which is much simpler than Eq.\ref{k0iyjy} and \ref{pmk0y} for the $ U(1)_{soc} $ case.
   Of course, this mapping only works near the QCP $ h \sim h_c $, will break down near $ h \sim 0 $ ( Fig.1c ).

   The effective action consistent with the $ U(1)_s $ symmetry and the translational symmetries is:
\begin{align}
	\mathcal{L}_{U(1)_s}= \psi^*\partial_\tau\psi+v^2|\nabla\psi|^2-\mu|\psi|^2+U|\psi|^4+ \cdots
\label{z2}
\end{align}
   where $ \mu= h-h_c$. It belongs to $z=2$ zero density SF-Mott transition universality class, therefore confirm the assumption used in \cite{z2}.

When $\mu<0$, $ \langle \psi  \rangle =0$ the mean field ground state is Z-FM state ( Fig.\ref{AFMh0} ).

While $\mu>0$, $ \langle  \psi  \rangle =me^{i\phi_0}$ where $ m= \sqrt{\mu/2U} $, the mean field ground state is
$$\mathbf{S}_i
=((-1)^{i_x+i_y}m\cos\phi_0,
(-1)^{i_x+i_y}m\sin\phi_0,
\sqrt{S^2-m})$$
 which is the canted co-planar AFM state ( Fig.\ref{AFMh}c ). It
 supports one gapless Goldstone mode $ \omega=ck $ due to the $ U(1)_s $ symmetry breaking.

   As stressed in the main text, $ U $ in Eq.\ref{z2} is the well known type-I dangerously irrelevant operator.
   In fact, it is also marginally irrelevant at the up-critical dimension $ d=2 $, but it is crucial to determine the
   symmetry breaking ground state and lead to the violation of the hyper-scaling at or above the upper critical dimension.

   When comparing with Eq.\ref{z2single}, one can see it is identical to Eq.\ref{z2single} near $ h= h_{c1} $ at the mirror symmetric point $ \beta=\pi/4 $.
   However, due to the dramatic difference between the  Eq.\ref{u1s} and Eq.\ref{k0iyjy} which express the {\sl very different}
   quantum spin order in terms of the {\sl identical } order parameter, the resulting symmetry breaking state
   the canted AFM in the former is completely different than the IC-SkX in the latter, but the Goldstone mode still takes the same form.
   However, away from the mirror symmetric point $ \beta \neq \pi/4 $, the DIO $ V $ term in Eq.\ref{z2single} moves in which does not touch the
   ground state, but changes its excitation to the exotic form Eq.\ref{Goldhc1}.

   Furthermore, in the SOC case studied in the main text, there are two critical fields $ h_{c1}, h_{c2} $ and an intermediate phase
   IC-SkX between the two critical ones. The Z-FM when $ h > h_{c1} $ is just a classical ground state instead of an exact ground state,
   so it supports strong quantum fluctuations, in sharp contrast to the Z-FM in Eq.\ref{u1s} which is an exact ground state with no quantum fluctuations.
   Here without the SOC, there is only one $ h_c $ which is very similar to $ h_{c1} $ at the Mirror symmetric point
   $ \beta=\pi/4 $, no intermediate phase ( Fig.\ref{AFMh0} ).
   These dramatic differences demonstrate the new features of the SOC in magnetic systems.


\begin{thebibliography}{99}
\bibitem{sameasEq10} Its symmetry breaking pattern and associated Goldstone mode is identical to Eq.\ref{coset}.
\bibitem{twoterms} In Eq.\ref{rhh}, the $ S^+ S^+ $ and $ S^- S^- $ terms
        belong to the Kitaev-like ( compass ) term.
        The third term consist of the term $ \cos 2 \beta ( S^{+}_i S^{-}_{i+y} + S^{-}_i S^{+}_{i+y} ) $ which belongs to
        the Heisenberg interaction, and also the term $  i \sin 2 \beta ( S^{+}_i S^{-}_{i+y} - S^{-}_i S^{+}_{i+y} ) $
        which belongs to the Dzyaloshinskii-Moriya (DM) interaction.
        This is due to the fact that Eq.1  contains the Heisenberg, Kitaev ( or compass ) and DM interaction as stressed in \cite{rh}.
\bibitem{type2}
In a classical or quantum phase transition, a dangerously irrelevant operator
is  irrelevant near the transition, but determines the nature of the symmetry breaking ground state.
We name the known one \cite{z2,rev1,scaling} as Type-I.
Here, we identify a new type of dangerously irrelevant operator which is irrelevant
near a QPT, does not change the ground state in the symmetry breaking phase, but changes its excitation spectrum to an exotic form.
We name the new one Type-II dangerously irrelevant operator (DIO).
In terms of" danger ", the type-II is weaker than type-I.
\bibitem{despite} Despite the universality class is completely determined by the order parameter $ \psi $, due to the
relation Eq.\ref{ordercs} between the quantum spin on the lattice and the order parameter $ \psi $ in the effective action,
the two states are very much different than those in the SF-Mott transition.
The boson BEC leads to a superfluid. While the mangnon BEC leads to a quantum magnetic order.
\bibitem{middle} Here we only focus on the the middle range $\beta_1<\beta<\beta_2$ in Fig.1.
\bibitem{not} Here $ u $ and $ v $ are two Type-II DIOs, should not be confused with the Bogliubov transformation coefficients $ u $ and $ v $ inEq.\ref{pmk0y}.
\end{thebibliography}
\end{document}